\documentclass[a4paper,11pt]{article}
\usepackage{jcappub}
\usepackage{amssymb}
\usepackage{graphicx}
\usepackage{amsmath}
\usepackage{hyperref}
\usepackage[defaultcolor=red]{changes}

\title{\boldmath Primordial black holes and induced gravitational waves from double-pole inflation}

\author[1]{Chengjie Fu,}
\author[2]{Shao-Jiang Wang$^*$}

\affiliation[1]{Department of Physics, Anhui Normal University, Wuhu, Anhui 241000, China}
\affiliation[2]{CAS Key Laboratory of Theoretical Physics, Institute of Theoretical Physics, Chinese Academy of Sciences, Beijing 100190, China}

\emailAdd{fucj@ahnu.edu.cn}
\emailAdd{schwang@itp.ac.cn (corresponding author)}

\abstract{The primordial black hole (PBH) productions from the inflationary potential with an inflection point usually rely heavily on the fine-tuning of the model parameters. We propose in this work a new kind of the $\alpha$-attractor inflation with asymmetric double poles that naturally and easily lead to a period of non-attractor inflation, during which the PBH productions are guaranteed with less fine-tuning the model parameters. This double-pole inflation can be tested against the observational data in the future with rich phenomenological signatures: (1) the enhanced curvature perturbations at small scales admit a distinctive feature of ultraviolet oscillations in the power spectrum; (2) the quasi-monochromatic mass function of the produced PBHs can be made compatible to the asteroid-mass PBHs as the dominant dark matter component, the planet-mass PBHs as the OGLE ultrashort-timescale microlensing events, and the solar-mass PBHs as the LIGO-Virgo events; (3) the induced gravitational waves can be detected by the gravitational-wave detectors in space and Pulsar Timing Array/Square Kilometer Array.}

\begin{document}
\maketitle
\flushbottom

\section{Introduction}

The formation of black holes could be not only of the astrophysical type but also of the primordial origin, the latter of which has recently attracted substantial attention from both observational and theoretical sides. From the observational ground, the massive primordial black holes (PBHs) could seed the supermassive black holes~\cite{Kawasaki:2012kn,Nakama:2017xvq} or even the stupendously large black holes in the galactic nuclei~\cite{Carr:2020erq} and hence the galaxy formation~\cite{Carr:2018rid}, and the solar-mass PBHs have been long conjectured as a possible explanation for the LIGO-Virgo events~\cite{Bird:2016dcv,Clesse:2016vqa,Sasaki:2016jop}, while the planet-mass PBHs could play a role in the ultra-short timescale microlensing events~\cite{2017Natur.548..183M,Niikura:2019kqi} or even the planet 9~\cite{Scholtz:2019csj}. Furthermore, the asteroid-mass PBHs are now the only open window for PBHs to account for all dark matter (DM)~\cite{Carr:2016drx,Katz:2018zrn,Niikura:2019kqi,Montero-Camacho:2019jte,Sugiyama:2019dgt} that is also consistent with FRB observations~\cite{Kainulainen:2021rbg}.

From the theoretical ground, PBHs could be produced by the re-entry of the enhanced curvature perturbations at small scales from the single-field ultra-slow-roll (USR) inflation~\cite{Garcia-Bellido:2017mdw,Ezquiaga:2017fvi,Germani:2017bcs,Motohashi:2017kbs,Kannike:2017bxn,Ballesteros:2017fsr,Cheng:2018yyr,Ozsoy:2018flq,Dalianis:2018frf,Fu:2019ttf,Mishra:2019pzq,Balaji:2022rsy,Ahmed:2021ucx,Karam:2022nym} (see also~\cite{Kawai:2021edk} for a dynamical brake) and multi-field curvaton model~\cite{Kawasaki:2012wr,Kohri:2012yw,Bugaev:2013vba,Ando:2017veq,Pi:2021dft,Maeso:2021xvl} or multi-field inflation with a second flat trajectory~\cite{Garcia-Bellido:1996mdl,Kawasaki:1997ju,Frampton:2010sw,Clesse:2015wea,Inomata:2017okj,Pi:2017gih,Inomata:2018cht,Cheong:2019vzl,Kawasaki:2019hvt,Palma:2020ejf,Fumagalli:2020adf,Braglia:2020eai,Anguelova:2020nzl,Gundhi:2020zvb,Gundhi:2020kzm,Kawai:2022emp,Balaji:2022dbi} as well as the resonance effects~\cite{Cai:2018tuh,Chen:2020uhe,Cai:2019jah,Cai:2019bmk,Cai:2021wzd,Cai:2021yvq} or even the modified sixth order dispersion relation~\cite{Ashoorioon:2018uey,Ashoorioon:2019xqc}. Another important channel comes from the collapse of various topological defects like primordial bubbles~\cite{Deng:2017uwc,Deng:2018cxb,Deng:2020mds,Ashoorioon:2020hln}, domain walls~\cite{Deng:2016vzb,Liu:2019lul}, oscillons~\cite{Cotner:2016cvr,Cotner:2017tir,Cotner:2018vug,Cotner:2019ykd}, and the delayed-decay false-vacuum regions in general first-order phase transitions~\cite{Liu:2021svg,Hashino:2021qoq,Liu:2022lvz,He:2022amv}  (see also~\cite{Baker:2021nyl,Baker:2021sno,Kawana:2021tde,Huang:2022him,Marfatia:2021hcp} for other similar mechanisms but with more specific model buildings). In particular,  PBH formations from the re-entry of the enhanced curvature perturbations at small scales could also induce secondary gravitational waves (GWs) detectable in the space-borne GW detectors and the Pulsar Timing Array (PTA) or Square Kilometer Array (SKA). See~\cite{Cai:2017cbj,Bian:2021ini} for recent reviews and references therein.

However, the inflationary model buildings with the appearance of a USR phase usually surfer from fine-tuning the model parameters, let alone to produce a significant amount of PBHs within a certain mass range. An intriguing approach~\cite{Kallosh:2000ve,Ferrara:2010yw,Ferrara:2010in,Kallosh:2013pby,Kallosh:2013lkr,Kallosh:2013hoa,Ferrara:2013rsa,Kallosh:2013daa,Kallosh:2013yoa,Galante:2014ifa,Linde:2015uga,Kallosh:2021mnu} to render a usual plateau in the inflationary potential is the introduction of a pole in the kinetic term, which, after transformed into the canonical form, would stretch the potential at the pole into the infinity in the field space so that an asymptotically flat potential could naturally emerge (see also~\cite{Hehl:1994ue,Gronwald:1997bx,Enckell:2018hmo,Gialamas:2019nly,Ghilencea:2020piz,Cai:2021png,Mikura:2020qhc,Mikura:2021ldx} for similar prospects from the Palatini formalism). In order to further generate a second extremely flat plateau near the end of the inflation that sufficiently decelerates the inflaton field, some elaborated polynomial potential~\cite{Dalianis:2018frf} or deformed Starobinsky potential~\cite{Mahbub:2019uhl,Mahbub:2021qeo} are invoked in the $\alpha$-attractor model with some delicate conditions to maintain an inflection point. It would be theoretically more appealing to make the inflaton velocity fall off exponentially without fine-tuning the model parameters in the potential.

In this paper, we propose a new model of $\alpha$-attractor inflation with double poles in its kinetic term, at which the inflationary potential in terms of the canonically normalized field are asymmetric (non-degenerate) due to the appearance of a nonzero  vacuum-expectation-value (vev) in the original potential. In this asymmetric double-pole inflation, the inflation with a rapidly-diminishing inflaton velocity can be more naturally and easily realized for a rather loose choice of values of model parameters so that PBH productions in our model require less fine-tuning. Nevertheless, some fine-tuning is still needed in order to generate the PBHs in the given mass range and abundance of observational interests, which is unharmful for our purpose.

\section{Model}

\begin{figure}
	\centering
	\includegraphics[width=0.9\columnwidth ]{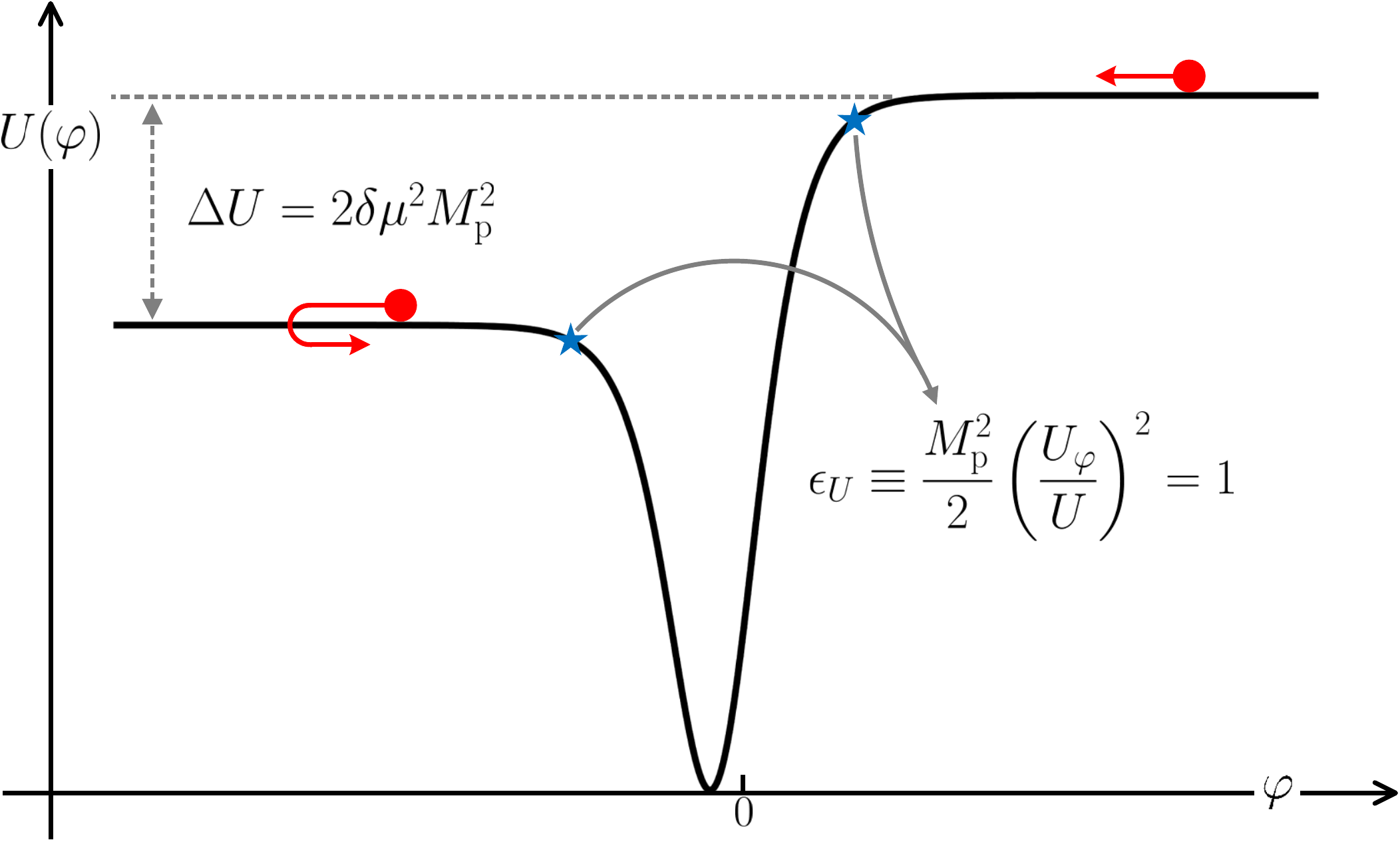}
	\caption{\label{fig1}  Schematic diagram of the canonicalised potential~\eqref{potential}.}
\end{figure}

The Lagrangian for the $\alpha$-attractor T-models \cite{Kallosh:2013hoa,Kallosh:2021mnu} is given by
\begin{align}
\mathcal{L} = \sqrt{-g}\left[ \frac{M_\mathrm{Pl}^2}{2}R - \frac{1}{2}\frac{\partial_\mu\phi\partial^\mu\phi}{\left(1-\phi^2/(6\alpha)\right)^2} - V(\phi) \right],
\end{align}
where $M_\mathrm{Pl}\equiv1/\sqrt{8\pi G}$ denotes the reduced Planck mass and $\alpha$ is a constant having a dimension of mass squared. The kinetic term of the inflaton field $\phi$ admits two poles at $\phi=\pm \sqrt{6\alpha}$, which can be canonically normalized in terms of a field,
\begin{align}\label{transformation}
\varphi = \sqrt{6\alpha} \tanh^{-1}\left(\frac{\phi}{\sqrt{6\alpha}}\right).
\end{align} 
Then, we consider the simple quadratic potential, however, with a nonzero vev $v>0$,
\begin{align}
V(\phi) = \frac{1}{2}m^2(\phi+v)^2.
\end{align}
Finally, the inflationary potential in terms of the canonically normalized field reads
\begin{align}\label{potential}
V(\phi(\varphi)) = \frac{1}{2}\mu^2 M_\mathrm{Pl}^2 \left( \tanh\frac{\varphi}{M} + \delta  \right)^2\equiv U(\varphi)
\end{align}
with $M\equiv \sqrt{6\alpha}$, $\mu \equiv mM/M_\mathrm{Pl}$, and $\delta\equiv v/M$. In Fig.~\ref{fig1}, this potential $U(\varphi)$ with $0<\delta<1$ is schematically illustrated with an intriguing potential difference $\Delta U=2\delta\mu^2M_\mathrm{Pl}^2$ between two asymptotically flat regions, where the potential slow-roll parameter $\epsilon_U\equiv M_\mathrm{Pl}^2(U_{\varphi}/U)^2/2 < 1$ on both ends. Consider the inflaton $\varphi$ slowly rolls down the potential from the right plateau to the left, leading to the standard single-field slow-roll inflationary phase until $\epsilon_U=1$. As long as the potential difference $\Delta U$ is sufficiently large, it may rapidly pass through the potential valley without stopping and then climb up the potential hill to the left plateau, resulting in a non-attractor inflation with a drastically decreasing field velocity until its kinetic energy fading away. After that, the inflaton field would turn back and then experience a second slow-roll phase, eventually settling down on the potential minimum. This dynamical picture is similar to those in Refs.~\cite{Saito:2008em,Fu:2020lob}. This paper implements the further exploration with physical motivation based on the previous toy model proposed in~\cite{Fu:2020lob}, which requires less fine-tuning in model parameters relative to the literature~\cite{Saito:2008em} but is purely phenomenological.

\begin{figure}
	\centering
	\includegraphics[width=0.9\columnwidth ]{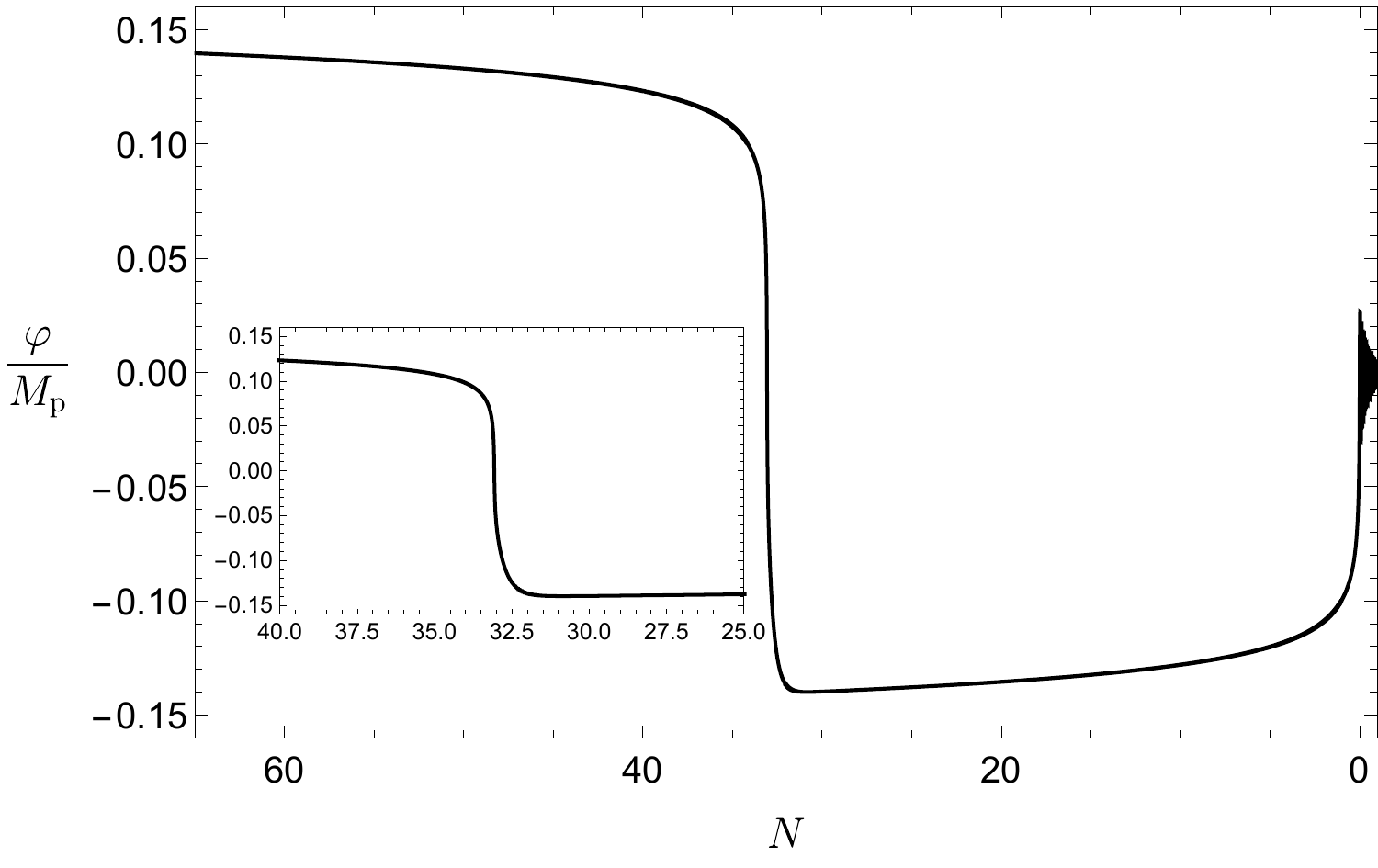}
	\caption{\label{fig2} The evolution of the inflaton $\varphi$ as a function of e-folding number $N \equiv \log(a_e/a)$, where $a_e$ denotes the scale factor at the end of second slow-roll inflation.}
\end{figure}

\begin{figure}
	\centering
	\includegraphics[width=1.\columnwidth ]{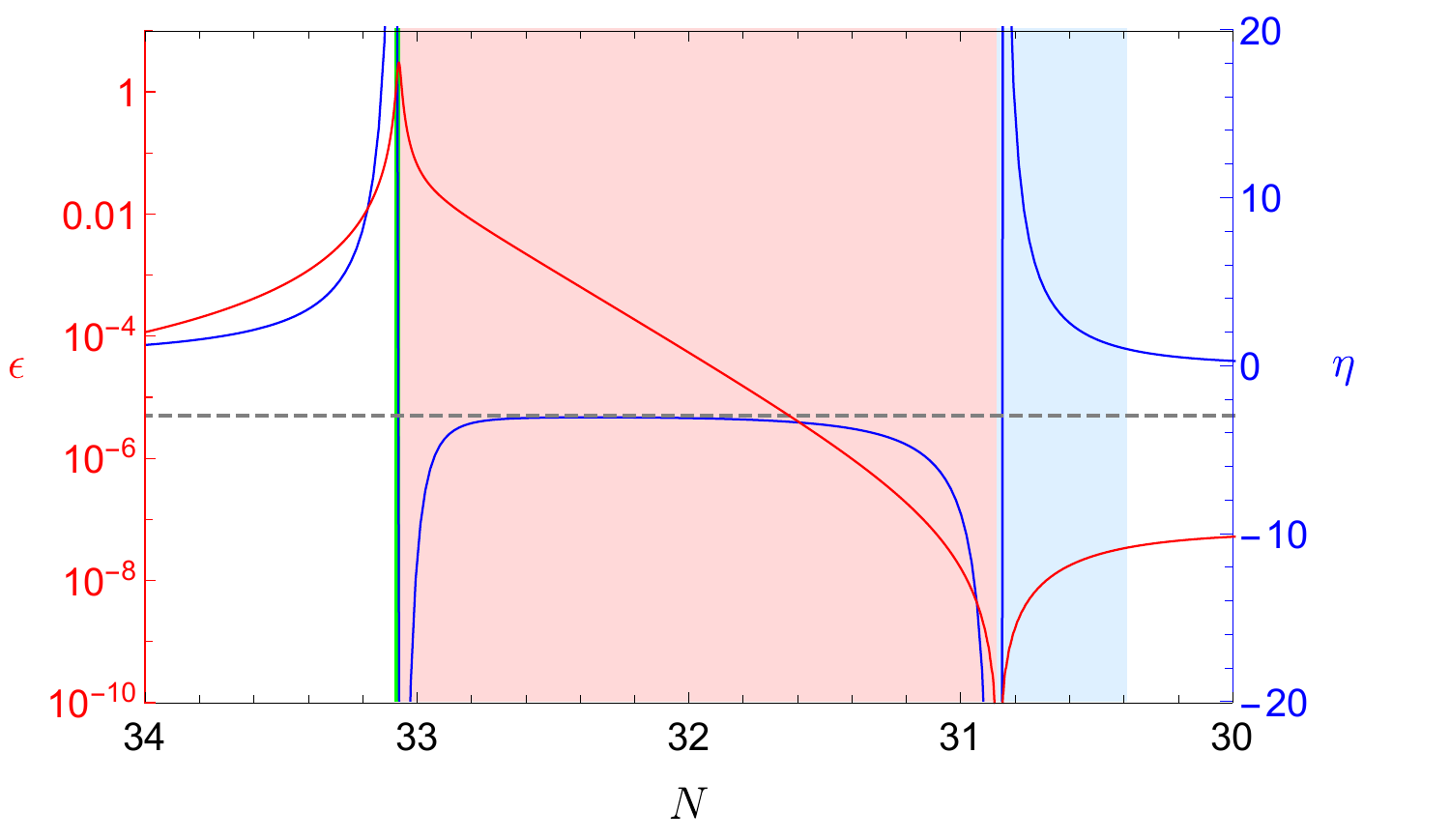}
	\caption{\label{fig3} The evolution of the slow-roll parameters $\epsilon$ (red line) and $\eta$ (blue line) as a function of e-folding number $N$.}
\end{figure}

For a canonical inflaton in the Einstein gravity, the curvature perturbation $\mathcal{R}$ in the momentum space obeys the following equation of motion,
\begin{align}\label{EoM_R}
\ddot{\mathcal{ R}}_k +\left(3+ 2\epsilon+2\eta\right)H\dot{\mathcal{ R}}_k+ \frac{k^2}{a^2}\mathcal{R}_k = 0,
\end{align}
with the Hubble slow-roll parameters $\epsilon \equiv -\dot H/H^2$ and $\eta \equiv \ddot\phi/H\dot\phi$. During the period of the non-attractor inflationary phase where the slow-roll condition breaks down by $\eta < - 3/2 $ when the inflaton climbs up the shallower plateau, the friction term in Eq.~\eqref{EoM_R} turns into a driving term, leading to the enhancement for the modes that exit the horizon around this phase. As a result, the curvature perturbations will exhibit a large bump in the power spectrum. For our model, the peak amplitude of the curvature power spectrum is mainly determined by the parameter $M$ with a negative correlation. It has been found that by setting $M \ll M_\mathrm{Pl}$, there is a significant amplification in the curvature power spectrum. If we fix the value of $M$, the parameter $\delta$ primarily determines the duration of the final slow-roll phase following the non-attractor inflation, and then controls the peak position of the curvature power spectrum. For a given $M$ value with $M \ll M_{\rm Pl}$, the parameter $\delta$ is constrained within a certain range around $(1.9\sim 2)M/M_{\rm Pl}$ to yield an appropriate duration of the final slow-roll phase. On the whole, the parameter choice in our case can be relatively loose in order to achieve the aforementioned inflationary dynamics and hence the significant amplification of the curvature perturbations.

Next, we exemplify with the following parameter set,
\begin{align}
 M/M_\mathrm{Pl} = 0.02, \qquad \delta=0.04,
\end{align}
to illustrate the background evolution of the inflaton $\varphi$ as a function of e-folding number $N$ as shown in Fig. \ref{fig2}.  We can see that the inflaton experiences a second period of the slow-roll evolution after turning back from the shallower plateau. Then in Fig.~\ref{fig3}, we plot the evolution for $\epsilon$ and $\eta$ as a function of e-folding number $N$. According to their evolution characteristics, we can split the dynamics of this model into five phases: 
\begin{enumerate}
\item\textit{Slow-roll phase}. Initially, the inflaton slowly rolls down along the right higher plateau of the potential. As the inflaton velocity gradually increases, the first slow-roll phase concludes at $N \simeq 33.078$ when $\epsilon=1$; 
\item\textit{Non-inflationary phase}. When the inflaton accelerates rapidly through the minimum of its potential, the Universe undergoes a momentary phase corresponding to the green-shaded region, which is non-inflationary and generate just $\Delta N \simeq 0.023$ \textit{e}-folds; 
\item\textit{Non-attractor phase}. Upon encountering the shallower plateau to the left, the inflaton climbs up the potential hill with a rapidly-diminishing velocity and eventually reaches the left plateau of the potential. Before it loses all its kinetic energy and turns back, the inflaton velocity decreases exponentially, resulting in a non-attractor phase with $\epsilon < 1$ and $\eta < -3/2$ for a short period. This phase, shown with the light red-shaded region, involves a brief quasi-constant-roll process with $\eta \simeq -3$ sandwiched by two shorter processes featuring a sharp dip in the $\eta$ evolution, and lasts for $\Delta N \simeq 2.194$ \textit{e}-folds; 
\item\textit{Over-damping inflationary phase}. When the inflaton just turns around, the inflation is not an attractor solution. During this phase, corresponding to the light blue-shaded region with $\Delta N \simeq 0.466$, the $\eta$ evolution exhibits a large sharp peak, leading a over-large friction term in Eq. \eqref{EoM_R}; 
\item\textit{Slow-roll phase}. Following the over-damping inflation, the Universe enters the slow-roll phase again, i.e. the second slow-roll inflation, generating about $\Delta N \simeq 30.395$ \textit{e}-folds. This unusual dynamical evolution leads to rich phenomenological signatures as we will elaborate below. 
\end{enumerate}

\section{Phenomenology}

In this paper, we consider that the inflation has two distinctive stages, pre-inflation and new inflation, where the former one is responsible for the production of the CMB-scale perturbations and the later one is governed by the scalar field $\varphi$. Note that we do not specify the pre-inflation model but simply assume that its predictions for the primordial perturbations at CMB scales are consistent with the Planck observational results~\cite{Planck:2018jri}. Before proceeding further, we should determine the energy scale and duration of the new inflation. On the one hand, as the energy scale for the new inflation should be far lower than that for the pre-inflation, we consider the case that the Hubble parameter of the new inflation is two orders of magnitude lower than that of the pre-inflation, which can be easily fixed if we assume that the pre-inflation is the standard slow-roll inflation and take the Hubble slow-roll parameter during pre-inflation $\epsilon_{\rm pre} = 10^{-3}$ as a fiducial value. Therefore, the parameter $\mu$ that determines the energy scale of the new inflation can be calculated by the relation,
\begin{align}
H_i = 10^{-2}\sqrt{8\pi^2\epsilon_{\rm pre} \mathcal{P_R}(k_\ast)},
\end{align}
where $H_i$ is the initial Hubble parameter of the new inflation and $\mathcal{P_R}(k_\ast)$ represents the power spectrum for the curvature perturbations at the CMB pivot scale $k_\ast = 0.05{\rm Mpc}^{-1}$.
On the other hand, we set the $e$-folding number from the time when the scale $k_\ast$ exits the horizon to the end of the inflation as $60$, and assume that the new inflation generates the last $50$ $e$-folds.

\subsection{Primordial curvature perturbations}

\begin{figure}
	\centering
	\includegraphics[width=0.8\linewidth]{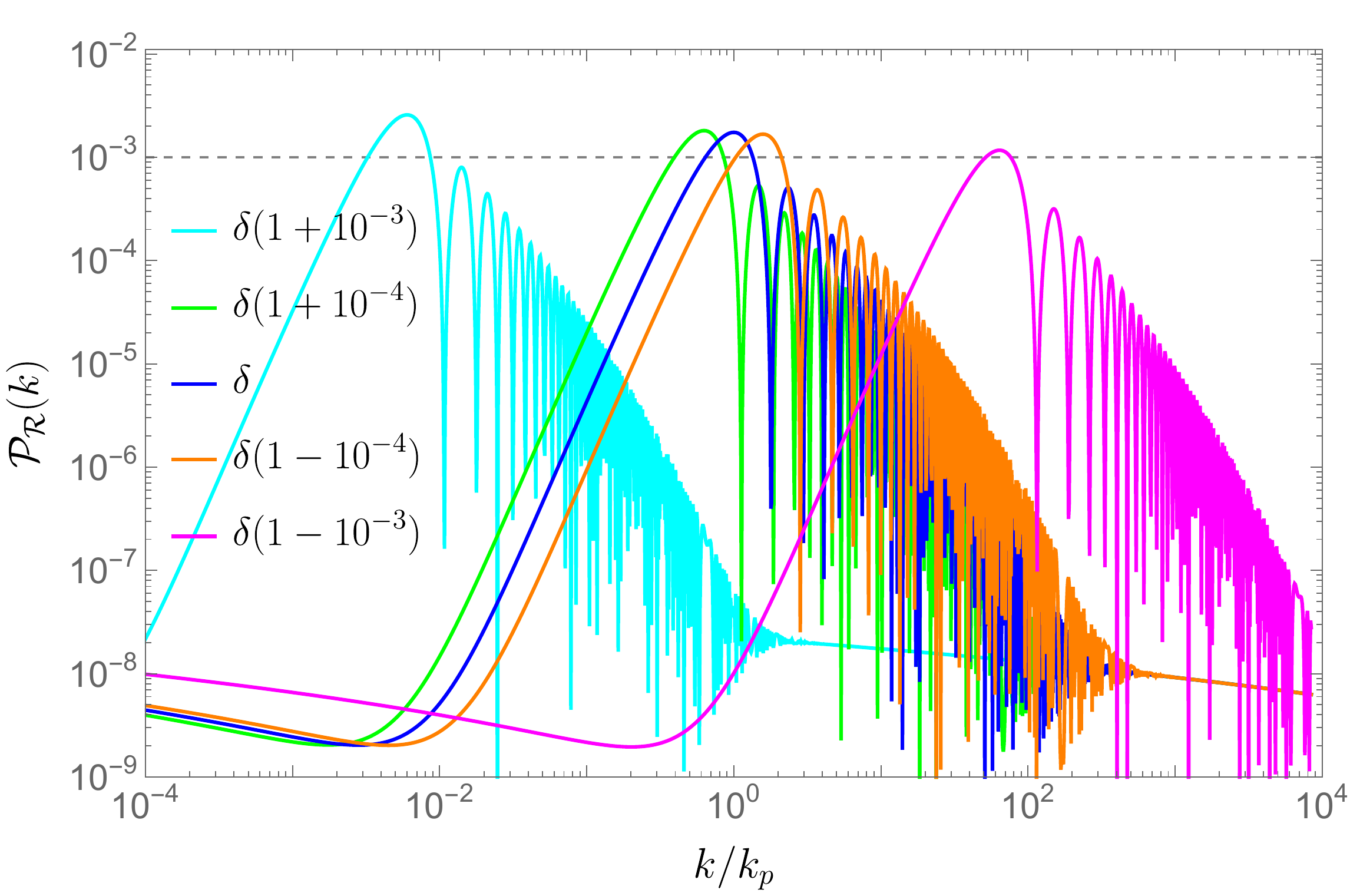}
	\caption{\label{fig4} The curvature power spectra for the fiducial parameter set (blue line) and the variations of the parameter $\delta$ by a factor of $1\pm10^{-4}$ and $1\pm10^{-3}$. Here, $k_p$ denotes the peak scale of the power spectrum for the fiducial parameter set.}
\end{figure}

\begin{table}[ht]
	\centering
	\begin{tabular}{cc|ccccc}
		\hline
		\hline
		Set  & &  & $M/M_\mathrm{Pl}$  & & &  $\delta$\\  
		\hline
		\hline
		\textit{1} & & & $0.0042$ & &  &$0.0081436$\\
		\textit{2} & & & $0.0046$ & & & $0.008936$ \\
		\textit{3} & & & $0.00475$ & & & $0.0092384$\\
		\hline
		\hline
	\end{tabular} 
	\caption{The successful parameter sets for producing some specific interesting populations of PBHs.}
	\label{table1}
\end{table}

\begin{figure}
	\centering
	\includegraphics[width=0.9\columnwidth ]{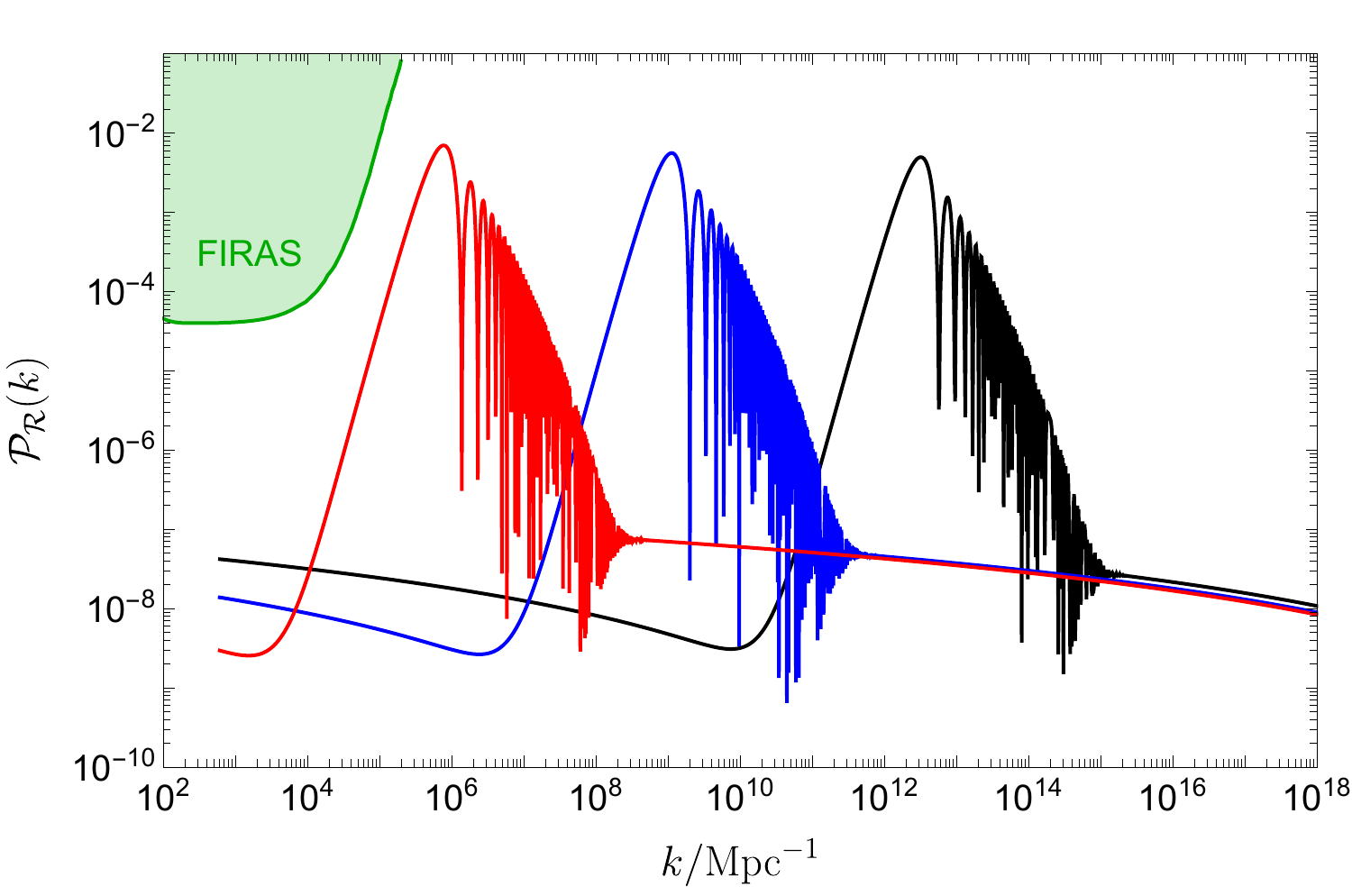}
	\caption{\label{fig5} The power spectra of the curvature perturbations for the parameter sets 1 (black), 2 (blue), and 3 (red) given in Table~\ref{table1}. Green-shaded region is excluded by the current constraint on $\mu$	distortion~\cite{Mather:1993ij,Fixsen:1996nj}. }
\end{figure}

The large peak in the power spectrum, typically $\mathcal{P_R} > \mathcal{O}(10^{-3})$, will lead to the production of a significant amount of PBHs. Before proceeding to adjust the parameters for generating PBHs of interest with appropriate abundance, it is crucial to examine the degree of fine-tuning required for PBH productions in relation to $\delta$. We choose $M=0.005M_{\rm Pl}$ and $\delta = 0.0097$ as a fiducial parameter set, which leads to a power spectrum with peak amplitude larger than $\mathcal{O}(10^{-3})$. In Fig. \ref{fig4}, we plot the curvature power spectra for the fiducial parameter set and the variations of $\delta$ by a factor of $1\pm10^{-4}$ and $1\pm10^{-3}$. The results indicate that shifts in $\delta$ by a factor of $1\pm10^{-4}$ have little impact on the peak position of the curvature power spectrum and a relatively minor effect on the peak amplitude of the curvature power spectrum. In the case of variations in $\delta$ by a factor of $1 \pm 10^{-3}$, there is a noticeable change in peak scale but the change in peak amplitude is still insignificant. The variations in $\delta$ by a factor of $1 \pm 10^{-3}$ still allow for a peak of magnitude higher than $\mathcal{O}(10^{-3})$. Compared to some other models \cite{Cole:2023wyx}, our scenario helps alleviate the fine-tuning of parameters to some extent for PBH productions. However, considering that the PBH abundance is exponentially sensitive to the power spectrum amplitude and that the peak position of the curvature power spectrum is highly susceptible to the variation in $\delta$, it is imperative to proceed with further fine-tuning to generate the PBHs in the given mass range and abundance of observational interests. In this work, we consider three parameter sets for successfully producing some specific interesting populations of PBHs as shown in Table~\ref{table1}, and the resulting power spectra of the curvature perturbations are shown in Fig.~\ref{fig5}. It is interesting to observe that the ultraviolet region of the bump in power spectrum displays an oscillating behavior, which is a distinctive feature of the curvature perturbations arising from the momentary non-inflationary phase followed by the non-attractor inflation. In the next subsection, we will calculate the mass and abundance of the produced PBHs for these three parameter sets.

\subsection{Primordial black holes}

\begin{figure}
	\centering
	\includegraphics[width=0.9\columnwidth]{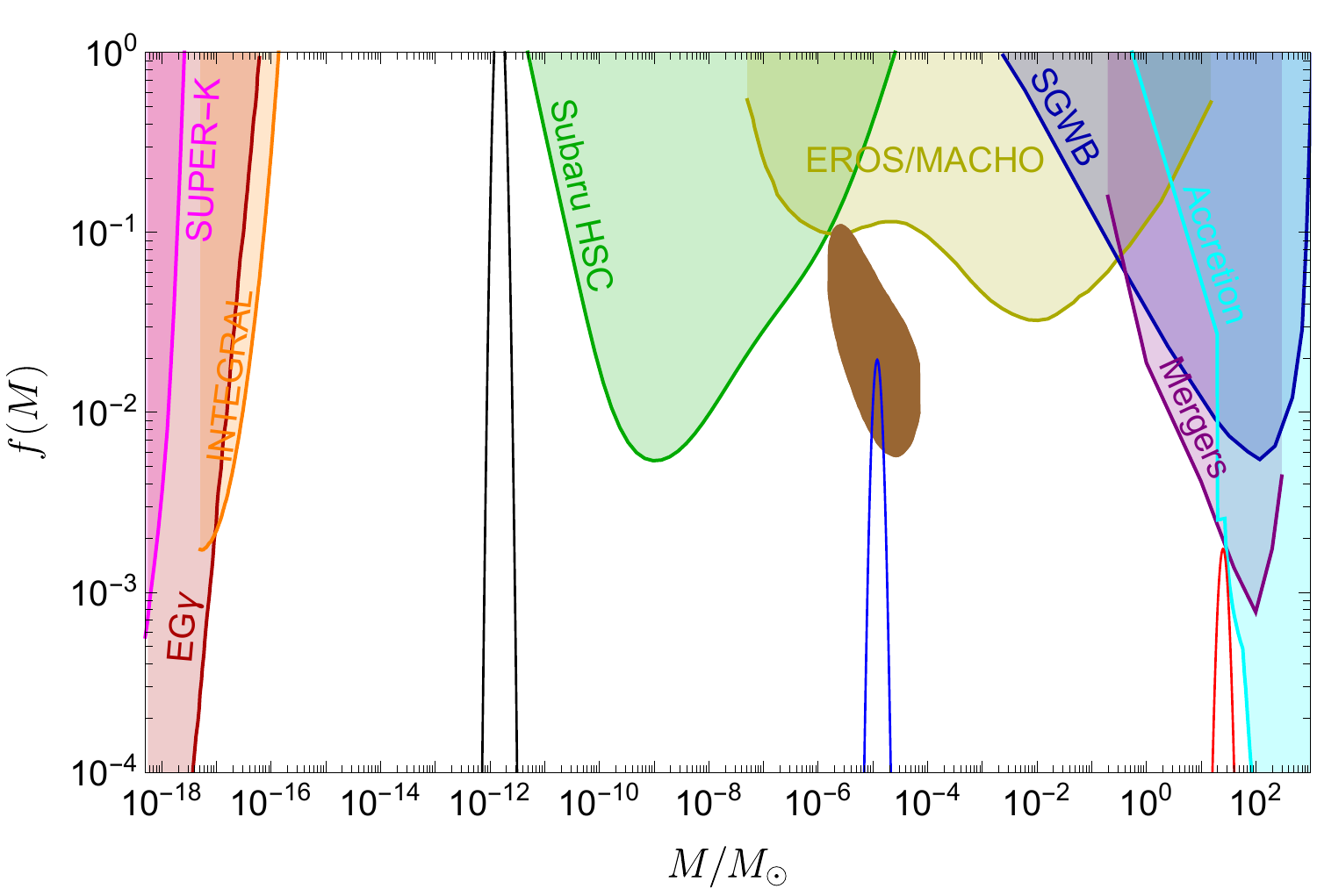}
	\caption{\label{fig6} The predicted PBH mass spectra for the parameter sets 1 (black), 2 (blue), and 3 (red) given in Table~\ref{table1}. The brown-shaded region represents the inferred PBH abundance by the OGLE ultrashort-timescale microlensing events~\cite{Niikura:2019kqi}, while the other shaded regions are current observational constraints on PBHs, which are summarized in the literature~\cite{Carr:2020gox,Green:2020jor} }
\end{figure}

The amplified curvature perturbations generated during the non-attractor phase of the inflation could eventually lead to the production of PBHs after their horizon reentering during the radiation-dominated era.  If these perturbations are Gaussian, the formation probability of PBHs on some smoothing comoving scale $R=k^{-1}$ is given according to the Press-Schechter theory by
\begin{align}
\beta(R) = \int_{\delta_c} \frac{d\delta}{\sqrt{2\pi\sigma^2(R)}}e^{-\frac{\delta^2}{2\sigma^2(R)}}\simeq \frac{1}{\sqrt{2\pi}\delta_c/\sigma(R)} e^{-\frac{\delta_c^2}{2\sigma^2(R)}},
\end{align}
where the threshold for PBH formations is usually taken as $\delta_c \simeq 0.45$ as suggested by several numerical studies~\cite{Musco:2004ak,Musco:2008hv,Musco:2012au}. Here, the variance of the smoothed density contrast, $\sigma^2(R)$, is related to the power spectrum of the curvature perturbation as~\cite{Blais:2002gw,Josan:2009qn}
\begin{align}
	\sigma^2(R) = \int_0^\infty d\ln{k} \widetilde{W}^2(R,k)\frac{16}{81}(kR)^4T^2(k,\tau=R)\mathcal{P_R}(k),
\end{align}
where $\tau$ denotes the conformal time and $T(k,\tau)$ is the scalar transfer function at the radiation-dominated era defined as
\begin{align}
	T(k,\tau) = 3 \frac{\sin(k\tau/\sqrt{3})-(k\tau/\sqrt{3})\cos(k\tau/\sqrt{3})}{(k\tau/\sqrt{3})^3}.
\end{align}
For the window function $\widetilde{W}$, we choose the real-space top-hat window function given by
\begin{align}
	\widetilde{W}(R,k) = 3\left( \frac{\sin(kR)-kR\cos(kR)}{(kR)^3} \right).
\end{align}
The mass of formed PBH is related to the smoothing scale $k^{-1}$ as~\cite{Ando:2018qdb}
\begin{align}
M(k) = M_\odot \left( \frac{\gamma}{0.2} \right) \left( \frac{g_\ast}{10.75} \right)^{-1/6} \left( \frac{k}{1.9\times 10^6 \; {\rm Mpc}^{-1}} \right)^{-2},
\end{align}
where $\gamma$ represents the collapsing efficiency and $g_\ast$ denotes the effective number of degrees of freedom for the energy density at PBH formation. In this paper, we take $\gamma\simeq0.2$ estimated with the simple analysis~\cite{Carr:1975qj} and adopt $g_\ast = 106.75$ as a fiducial value. The current mass spectrum of produced PBHs is given by~\cite{Sasaki:2018dmp}
\begin{align}
	f(M) &\equiv \frac{1}{\Omega_{\rm DM}} \frac{d \Omega_{\rm PBH}}{d\ln M} \nonumber\\
	    &= \frac{\beta(M)}{1.84\times10^{-8}} \left(\frac{\gamma}{0.2}\right)^{3/2}\left(\frac{g_{\ast}}{10.75}\right)^{-1/4}\left(\frac{M}{M_\odot}\right)^{-1/2},
\end{align}
where $\Omega_{\rm DM}$ represents the current DM density parameter. So, the total fraction of PBHs in DM can be estimated from
\begin{align}
\frac{\Omega_{\rm PBH}}{\Omega_{\rm DM}} = \int d\ln M\,f(M).
\end{align}
In Fig.~\ref{fig6}, we plot the current mass spectra of PBHs produced by the power spectra shown in Fig.~\ref{fig5}. One can see that each of the resulting PBH mass spectra admits a sharp peak, and they peak at PBH masses around $\mathcal{O}(10^{-12})M_\odot$, $\mathcal{O}(10^{-5})M_\odot$ and $\mathcal{O}(10)M_\odot$, respectively. These PBHs, comprising $\mathcal{O}(100)\%$ (set 1), $\mathcal{O}(1)\%$ (set 2), and $\mathcal{O}(0.1)\%$ (set 3) of total DM, respectively, can explain all DM, OGLE ultrashort-timescale microlensing events~\cite{Niikura:2019kqi}, and LIGO-Virgo GW events~\cite{Bird:2016dcv,Clesse:2016vqa,Sasaki:2016jop}, respectively.

\subsection{Scalar induced gravitational waves}

\begin{figure}
	\centering
	\includegraphics[width=0.9\columnwidth ]{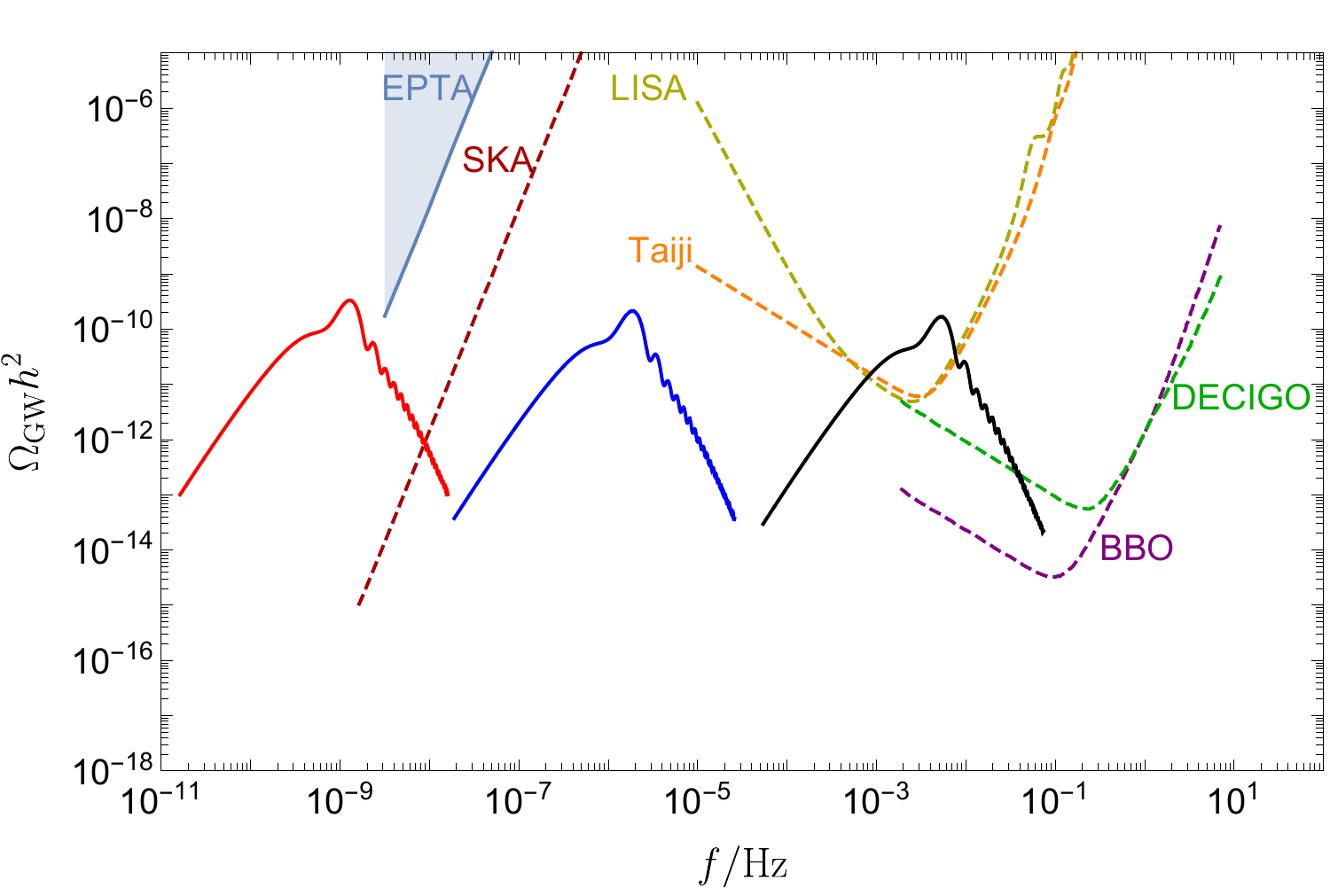}
	\caption{\label{fig7} The predicted current energy spectra of SIGWs for the parameter sets 1 (black), 2 (blue), and 3 (red) given in Table~\ref{table1}. The shaded region represents the current constraint by EPTA~\cite{Lentati:2015qwp}. The dashed lines are the expected sensitivity curves of the future GW observations, including SKA~\cite{Janssen:2014dka}, LISA~\cite{LISA:2017pwj}, Taiji~\cite{Ruan:2018tsw}, DECIGO~\cite{Kawamura:2011zz}, and BBO~\cite{Phinney:2003}.}
\end{figure}

During the radiation-dominated era, the enhanced curvature perturbations that produce a considerable amount of PBHs will inevitably induce a significant GW background according to the second-order cosmological perturbation theory\footnote{Note that we neglect possible effects of non-Gaussianities \cite{Atal:2021jyo} and one-loop corrections \cite{Chen:2022dah} to the induced GW background in present paper.} (see~\cite{Domenech:2021ztg} for a recent review and~\cite{Domenech:2019quo,Domenech:2020kqm}  for a general constant equation of state). The scalar-induced GWs (SIGWs) are generated mainly around horizon reentry, and stop growing as the scalar perturbations decay soon after horizon reentry. We define $\tau_c$ as the moment when the density ratio of GWs to the background radiation becomes a constant. At $\tau_c$, the density parameter of SIGWs per logarithmic interval of $k$ is calculated analytically as~\cite{Espinosa:2018eve,Kohri:2018awv}
\begin{align}
\Omega_{\rm{GW}}(k,\tau_c)& = \frac{1}{12} \int^\infty_0 dv \int^{|1+v|}_{|1-v|}du \left( \frac{4v^2-(1+v^2-u^2)^2}{4uv}\right)^2 \nonumber \\
& \times \mathcal{P}_\mathcal{R}(ku)\mathcal{P}_\mathcal{R}(kv)\left( \frac{3}{4u^3v^3}\right)^2 (u^2+v^2-3)^2 \nonumber\\ 
& \times \left\{\left[-4uv+(u^2+v^2-3) \ln\left| \frac{3-(u+v)^2}{3-(u-v)^2}\right| \right]^2 \right.  \nonumber \\
& \left. + \pi^2(u^2+v^2-3)^2\Theta(v+u-\sqrt{3})\right\}.
\end{align}
The energy spectrum observed today for SIGWs is given by~\cite{Ando:2018qdb}
\begin{align}
\Omega_{\rm{GW}}(k) = 0.83\left( \frac{g_\ast}{10.75} \right)^{-1/3}\Omega_{\rm{r},0}\Omega_{\rm{GW}}(k,\tau_c)\;,
\end{align}
where $\Omega_{\rm{r},0}$ is the current density parameter of radiation. The energy spectra of SIGWs predicted by our model are presented in Fig.~\ref{fig7}. It is intriguing to
observe that the GW energy spectrum exhibits an oscillating structure in the ultraviolet region, and a similar feature can be found in the previous toy model~\cite{Fu:2020lob} about this type of inflationary dynamics.
For the parameter set 1, the predicted GW signal is located in the frequency range of the space-based GW detectors, e.g. LISA, Taiji, deci-hertz interferometer GW observatory (DECIGO), and big bang observer (BBO), and can be tested by these GW observations. In the case of adopting the parameter set 3, while the predicted GW signal evades the current constraint from EPTA, its energy spectrum exceeds the sensitivity of SKA. However, SIGWs for parameter set 2 can not be probed by future GW observations. Finally, it should be noted that the estimation of PBH abundance involves uncertainties in the choice of a window function~\cite{Ando:2018qdb}. If we take the Gaussian window function, the larger curvature perturbations are required for the same PBH abundance compared to the window function we used in this paper. In this case, the infrared region of the resulting GW spectrum for set 2 can exceed the sensitivity curve of SKA, but the resulting GW spectrum for set 3 will fail to avoid the current EPTA constraint.

\section{Conclusion}

The PBH productions from the current inflation models with enhanced curvature perturbations at small scales are not always guaranteed as it usually elaborates on some delicate conditions to generate a second extremely flat plateau near the end of the inflation. We propose a new kind of the $\alpha$-attractor inflation model that admits double poles in its kinetic term and a nonzero vev in its potential term. The resulting inflationary potential is asymmetric (non-degenerate) at these two poles, ensuring the existence of a period of non-attractor inflation for the PBH productions. The associated phenomenological signatures are also predicted with parameter choices of observational interest for future detections. Our proposal is certainly not limited to the current model building as it can be easily generalized into other forms sharing the same feature with double poles and a nonzero vev in its kinetic and potential terms, respectively. A pursuit for its ultraviolet completion from the superconformal constructions is also theoretically desirable in the future to reveal the origin of the nonzero vev.

In conclusion, we provide here a brief discussion concerning the one-loop corrections in the curvature power spectrum, which have been widely discussed recently in the USR inflation \cite{Kristiano:2022maq,Kristiano:2023scm} and the resonance model with the oscillatory feature in potential \cite{Inomata:2022yte}. More specifically, it was argued in Refs.~\cite{Kristiano:2022maq,Kristiano:2023scm} that the amplified small-scale modes producing significant amount of PBHs generically induce large one-loop corrections to the CMB-scale curvature power spectrum. The authors concluded that PBH formation based on a USR phase in single-field inflation is not viable, which was criticized recently in Refs.~\cite{Riotto:2023hoz,Riotto:2023gpm} (see also~\cite{Choudhury:2023vuj,Choudhury:2023jlt,Choudhury:2023rks,Choudhury:2023hvf}). Ref.~\cite{Firouzjahi:2023aum} found that for an infinitely sharp transition from the USR phase into the final slow-roll phase, the induced one-loop corrections can be arbitrarily large, invalidating the perturbative approximation completely. However, if the transition occurs smoothly, the dangerous one-loop corrections are washed out during the subsequent evolution of the mode functions after the USR phase, thus supporting the arguments in \cite{Riotto:2023gpm}. A similar conclusion was reached more clearly in \cite{Firouzjahi:2023ahg}, where the one-loop corrections to the CMB-scale curvature power spectrum were calculated by using the $\delta N$ formalism. Moreover, Ref.~\cite{Motohashi:2023syh} provides a straightforward counterexample to the no-go theorem of PBH production from single-field inflation claimed in \cite{Kristiano:2022maq,Kristiano:2023scm} by considering the transient constant-roll inflation. Note that the situation in our model is somewhat different. Firstly, the evolution of the slow-roll paramters $\epsilon$ and $\eta$ markedly diverges from that found in realistic USR inflation, as seen from Fig. \ref{fig3}. Secondly, while previous calculations about one-loop corrections dismissed the higher-order contributions of $\epsilon$ in the interaction Hamiltonians, we cannot assume that $\epsilon$ is negligible in our model, as there exists a non-inflationary phase with $\epsilon > 1$. Consequently, we cannot hastily deduce whether the mechanism of PBH formation is effective in our scenario, and further numerical investigation is necessary to understand the details of the one-loop corrections in curvature power spectrum, which exceeds the scope of the current research.

\acknowledgments
This work is supported by the National Key Research and Development Program of China Grant  No. 2021YFC2203004, No. 2020YFC2201501 and No. 2021YFA0718304, 
the National Natural Science Foundation of China Grants No. 12105344, No. 12047503 and No.12235019, the Key Research Program of the Chinese Academy of Sciences (CAS) Grant No. XDPB15, the Key Research Program of Frontier Sciences of CAS, 
and the Science Research Grants from the China Manned Space Project with No. CMS-CSST-2021-B01.


\bibliographystyle{JHEP}
\bibliography{references}

\end{document}